\documentclass[11pt, a4paper, logo, copyright]{googledeepmind}

\usepackage[authoryear,round]{natbib}
\usepackage{placeins}
\usepackage{multirow}
\usepackage{wrapfig}
\usepackage{xurl}

\usepackage{amsmath,amsfonts,bm}

\def\eqref#1{equation~\ref{#1}}

\def\1{\bm{1}}

\DeclareMathAlphabet{\mathsfit}{\encodingdefault}{\sfdefault}{m}{sl}
\SetMathAlphabet{\mathsfit}{bold}{\encodingdefault}{\sfdefault}{bx}{n}

\newcommand{\modelname}{\textsc{AudioFM}}
\definecolor{evalgreen}{HTML}{267347}
\definecolor{evalred}{HTML}{B13A3A}
\newcommand{\evalcheck}{\textcolor{evalgreen}{\ding{51}}}
\newcommand{\evalcross}{\textcolor{evalred}{\ding{55}}}

\title{AudioFM: Towards General Purpose Audio Representation Learning}
\reportnumber{}
\author[1,3,*]{Yamin~Li}
\author[1]{Ruidong~Zhang}
\author[1]{Anran~Wang}
\author[2]{Eduardo~Fonseca}
\author[2]{Dan~Ellis}
\author[1]{Zechen~Li}
\author[1]{Simon~A.~Lee}
\author[1]{Tao~Pang}
\author[1]{Jake~Garrison}
\author[1]{Maxwell~A.~Xu}
\author[2]{Yicheng~Wang}
\author[1]{Jitian~Zhang}
\author[1]{Weizhi~Zhang}
\author[1]{Haichao~Zhang}
\author[3]{Catie~Chang}
\author[1,4]{Xin~Liu}
\author[1]{Anupam~Pathak}
\author[1,4]{Daniel~McDuff}
\author[1,4]{Shwetak~Patel}
\author[1,4]{Jake~Sunshine}
\author[1]{Yuzhe~Yang}
\author[1]{Xiaoran~Fan}

\affil[1]{Google Research}
\affil[2]{Google DeepMind}
\affil[3]{Vanderbilt University}
\affil[4]{University of Washington}
\affil[*]{Work done during an internship at Google}

\begin{document}

\begin{abstract}
Learning audio representations that effectively bridge holistic global soundscapes with fine-grained acoustic moments is a long-standing goal in self-supervised learning. Recording-level contrastive objectives encourage agreement between pooled audio representations, but do not explicitly supervise local time-frequency representations. We introduce \textbf{\modelname},
a self-supervised framework that combines recording-level contrast with patch-wise masked latent discrimination. Using visible acoustic context, a student predicts representations of masked patches and identifies their corresponding full-view teacher targets among candidates from other recordings. We further incorporate weaker supervision from eligible temporal neighbors to encourage local temporal consistency. Extensive evaluations show that \modelname{} achieves state-of-the-art performance across diverse audio domains in linear probing, fine-tuning, and zero-shot transfer, demonstrating superior event detection, robust scene-level understanding, and strong domain generalizability.
\end{abstract}

\maketitle

\section{Introduction}
Audio carries information beyond spoken language, offering insight into our surroundings, biological activity, and human states. Developing separate supervised models for these applications requires task-specific annotations, whose collection can be costly and often depends on domain expertise~\citep{zhang2024opera,schwinger2025biofoundation}. Audio self-supervised learning (SSL) seeks to learn general representations from unlabeled recordings that can be transferred across downstream tasks~\citep{huang2022audiomae,niizumi2024m2d,huang2026audiomosaic,chen2023beats,chen2024eat,alex2025sslam}. These methods provide a path toward audio foundation models that support diverse acoustic applications. Progress towards this goal depends on both effective self-supervised learning and systematic evidence of transfer across acoustic domains. Yet how consistently improvements on standard audio benchmarks extend to diverse specialized domains remains insufficiently understood. Understanding how performance changes with model and data scale further informs the development of these models.

From a modeling perspective, supporting diverse acoustic applications requires representations that preserve information at different levels of detail. A recording can contain sustained backgrounds, overlapping sources, and brief events, whose relevance depends on the task. Overall acoustic context can inform scene recognition, while subtle temporal and spectral differences can distinguish animal vocalizations or patterns in physiological sounds. Information that is secondary for one task may therefore be central to another. A general-purpose representation should capture the overall content of a recording while retaining the finer structure needed to recognize and locate its constituent events. Joint global and local supervision is important for capturing recording-level context alongside fine-grained acoustic content~\citep{chen2024eat,alex2025sslam}. A key modeling question is how to make individual patch representations discriminative while maintaining recording-level consistency.

Establishing the generality of these representations, in turn, requires examining their transfer across acoustic domains and task settings. Broad audio benchmarks such as HEAR~\citep{turian2022hear} and HARES~\citep{wang2021hares} already assess transfer across speech, music, and environmental sound, while more recent studies in health acoustics~\citep{zhang2024opera} and bioacoustics~\citep{schwinger2025biofoundation} have started to investigate specialized applications. However, differences in model coverage and evaluation protocols leave an incomplete picture of the relative strengths of audio SSL encoders across general and specialized acoustic domains. Alongside transfer, examining performance as model capacity and pretraining data increase can help assess the potential for further improvement.

We introduce \textbf{\modelname}, an audio foundation model trained through joint global-local discrimination. The model combines recording-level contrast between complementary time- and frequency-masked views with a local objective that discriminates full-view teacher targets for masked regions from targets in other recordings. We assess \modelname{} and existing audio SSL encoders through a shared linear-probing pipeline across a diverse range of audio tasks and domains (Table~\ref{tab:evaluation-coverage}). Complementary evaluations cover supervised fine-tuning, frame-level event detection, and transfer of source-trained classifiers to new datasets or speakers without target-domain training, alongside an analysis of how performance scales with model size and pretraining data volume.
We summarize our contributions as follows:
\begin{itemize}[leftmargin=*]
    \item \textbf{A general-purpose audio foundation model.} \modelname{} combines recording-level contrast with masked latent discrimination to learn an encoder for diverse acoustic tasks.

    \item \textbf{Systematic cross-domain evaluation.} We compare frozen audio encoders across 12 classification tasks in four acoustic domains using shared task-specific probing protocols.

    \item \textbf{Model and data scaling.} We observe clear and consistent performance gains as encoder capacity and pretraining data volume increase, characterizing empirical scaling trends for audio pretraining.

    \item \textbf{Strong downstream performance and transfer.} \modelname{} achieves state-of-the-art fine-tuning and temporal detection results while demonstrating strong generalizability to unseen data.
\end{itemize}

\begin{table}[t]
\centering
\begin{minipage}{\linewidth}
\centering
\caption{\textbf{Downstream evaluation coverage.}
\evalcheck/\evalcross: dedicated evaluation reported/not reported in the original study;
SED: frame-level sound event detection.
}
\label{tab:evaluation-coverage}
\begingroup
\fontsize{7.5}{9.0}\selectfont
\setlength{\tabcolsep}{1.8pt}
\renewcommand{\arraystretch}{1.05}
\begin{tabularx}{\linewidth}{@{}l*{6}{>{\centering\arraybackslash}X}@{}}
\toprule
\textbf{Study} & \shortstack{General\\audio} & \shortstack{Speech\\recog.} & Bioacoustics & Medical & \shortstack{Speech\\emotion} & SED \\
\midrule
AudioMAE~\citep{huang2022audiomae} & \evalcheck & \evalcheck & \evalcross & \evalcross & \evalcross & \evalcross \\
BEATs~\citep{chen2023beats} & \evalcheck & \evalcheck & \evalcross & \evalcross & \evalcheck & \evalcross \\
ATST-Frame~\citep{li2024atst} & \evalcheck & \evalcheck & \evalcross & \evalcross & \evalcross & \evalcheck \\
M2D~\citep{niizumi2024m2d} & \evalcheck & \evalcheck & \evalcross & \evalcheck & \evalcheck & \evalcross \\
EAT~\citep{chen2024eat} & \evalcheck & \evalcheck & \evalcross & \evalcross & \evalcross & \evalcross \\
SSLAM~\citep{alex2025sslam} & \evalcheck & \evalcheck & \evalcross & \evalcross & \evalcross & \evalcross \\
BAT~\citep{ghaffari2026bat} & \evalcheck & \evalcheck & \evalcheck & \evalcross & \evalcross & \evalcheck \\
AudioMosaic~\citep{huang2026audiomosaic} & \evalcheck & \evalcheck & \evalcross & \evalcross & \evalcross & \evalcross \\
\midrule
\rowcolor{black!5}
\textbf{\modelname{} (Ours)} & \evalcheck & \evalcheck & \evalcheck & \evalcheck & \evalcheck & \evalcheck \\
\bottomrule
\end{tabularx}
\endgroup
\end{minipage}
\end{table}

\section{Related Work}
\label{sec:related_work}

\paragraph{General-purpose audio representation learning.}
Audio pretraining seeks to provide a common representation for diverse downstream tasks.
Large-scale event classification and audio-text alignment offer supervised and language-supervised routes to transferable features~\citep{kong2020panns,elizalde2023clap}.
Within audio SSL, one line of work emphasizes agreement between different views of a recording, through contrastive learning or non-contrastive self-distillation~\citep{saeed2021cola,niizumi2021byola,huang2026audiomosaic}.
Another learns by predicting masked content, with the choice of target defining the supervision: spectrogram reconstruction operates on the acoustic signal~\citep{huang2022audiomae}, while acoustic token prediction uses learned discrete targets~\citep{chen2023beats}.
Continuous latent prediction instead uses teacher representations as targets, a recipe shared by most recent audio encoders~\citep{niizumi2024m2d,chen2024eat,alex2025sslam,ghaffari2026bat}.
These objectives emphasize different aspects of the audio signal, making downstream transfer an important test of which acoustic distinctions their representations preserve.

\paragraph{Foundation models and cross-domain evaluation.}
Beyond audio, foundation-model research has pursued broadly transferable representations for visual understanding~\citep{oquab2024dinov2}, general time-series analysis~\citep{goswami2024moment,timesfm2024}, and wearable and physiological applications~\citep{narayanswamy2025lsm,zhang2025sensorlm,li-etal-2025-sensorllm, li2026glucofm, lee2026himae, li2026mind, zhou2026physiologyaware, chen2026anymo, LiCha_CBrain_MICCAI2025}.
Across these settings, a central question is how effectively pretrained representations support different downstream tasks and data distributions.
For audio, this question spans both general audio understanding and specialized acoustic applications.
HEAR and HARES assess representations across speech, music, and environmental sound \citep{turian2022hear,wang2021hares}, while OPERA and BioFoundation provide domain-focused evaluations in respiratory acoustics and bioacoustics \citep{zhang2024opera,schwinger2025biofoundation}.
However, coverage of general and specialized acoustic domains remains distributed across separate evaluations. Meanwhile, comparisons based on published fine-tuning results can reflect differences in downstream training recipes as well as pretrained representations. We therefore complement such results with a shared frozen-encoder evaluation across general and specialized acoustic tasks.

\paragraph{Global-local supervision and masked discrimination.}
Global and local supervision address complementary requirements: consistency across views of an instance and sensitivity to its constituent acoustic content.
AudioMosaic studies recording-level contrast between complementary time- and frequency-masked views~\citep{huang2026audiomosaic}, while EAT and SSLAM explicitly couple utterance- and patch-level regression~\citep{chen2024eat,alex2025sslam}.
Joint global-local contrast also has precedents in vision \citep{wang2021densecl,zhou2022mimco}.
Beyond supervision granularity, the prediction targets and candidate sets determine what a local contrastive objective distinguishes.
SSAST and MAE-AST discriminate raw masked spectrogram patches using other masked positions in the same recording as candidates~\citep{gong2022ssast,baade2022maeast}; SPIRAL and ConMIM instead contrast learned teacher representations, with negatives drawn from other positions in the same utterance or image~\citep{huang2022spiral,yi2023conmim}.
\modelname{} combines recording-level view contrast with contextual prediction of full-view teacher features at masked locations, drawing competing patch targets from other recordings at the same target coordinate.
This candidate construction avoids requiring acoustically similar regions within a recording to be distinguished solely because they occupy different positions.
We also give weaker supervision to nearby teacher targets along the time axis when those positions are masked in the student, encouraging local temporal consistency within this discriminative objective.

\section{Method}
\label{sec:method}

\begin{figure}[t]
    \centering
    \includegraphics[width=\linewidth]{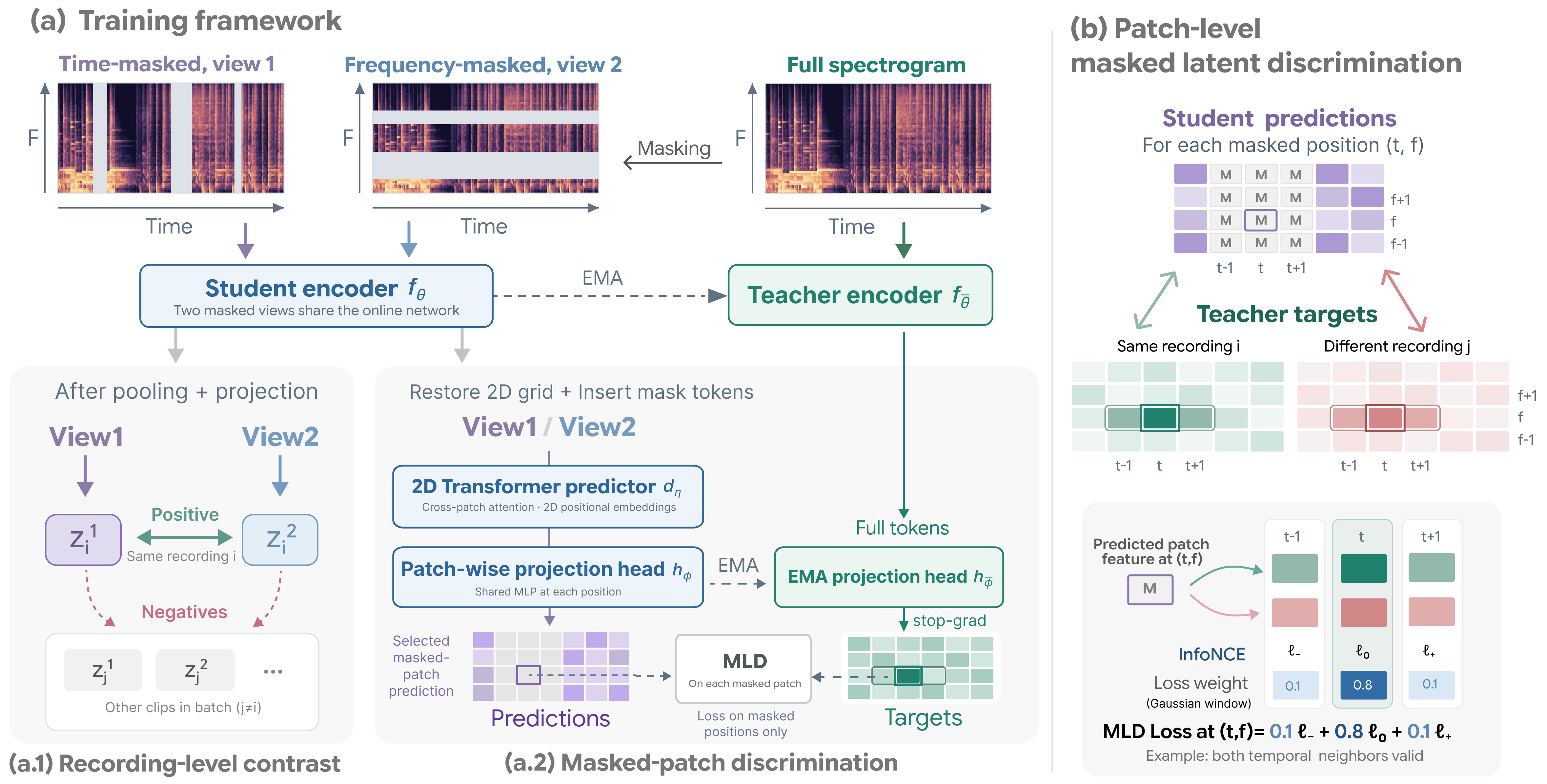}
    \caption{\textbf{Overview of \modelname{}.} (a) A shared student encoder learns from complementary time- and frequency-masked views through recording-level contrast (a.1) and masked-patch prediction (a.2). An EMA teacher encodes the corresponding unmasked views to provide latent targets. (b) Detailed view of masked-patch discrimination in (a.2). 
    }
    \label{fig:framework}
\end{figure}

\paragraph{Framework overview.}
\label{sec:framework} 
\modelname{} combines recording-level contrast across masked views with masked-patch latent prediction against full-view teacher targets (Figure~\ref{fig:framework}).
For each recording $i$ in a batch of size $B$,
we independently apply waveform augmentations to obtain
two versions of the recording, following
AudioMosaic~\citep{huang2026audiomosaic}.
We convert them into log-Mel spectrograms $X_i^1$ and $X_i^2$,
and embed each spectrogram as a $T \times F$ grid
of non-overlapping patch tokens, where $T$ and $F$ denote
the numbers of patches along the time and frequency axes,
respectively. Patch positions are indexed by
$\mathcal{P}=\{1,\ldots,T\}\times\{1,\ldots,F\}$.
We then apply complementary structured masking to these
tokens: view $1$ removes tokens at selected temporal
positions across all frequency bands, while view $2$
removes tokens at selected frequency bands across all
time steps.
Let $\mathcal{M}_i^v\subset\mathcal{P}$ denote the masked
positions in view $v\in\{1,2\}$.
A shared Vision Transformer student encoder $f_\theta$ processes the visible patches with their positional embeddings and a class token, while the teacher processes all patch tokens of the corresponding spectrogram.
From the encoded visible tokens, the student splits into two parallel pathways: a \textit{global branch} for recording-level contrast (Figure~\ref{fig:framework}(a.1)) and a \textit{local branch} for masked-patch discrimination (Figure~\ref{fig:framework}(a.2)). 

For global contrast, pooled encoder outputs pass through a shared projection head and $\ell_2$ normalization to produce recording embeddings $\mathbf{z}_i^v$. The positive pair consists of the two masked views of the same recording. 
For local prediction, we remove the class token and restore the visible patch representations at corresponding grid coordinates. We fill masked positions with a shared learnable mask token of dimension $D$.
A shared 2D Transformer predictor $d_\eta$ projects these features to a bottleneck dimension $D_{\mathrm{pred}}$, adds 2D sinusoidal positional encodings and a view-specific masking-branch embedding, applies full time-frequency self-attention, and projects the outputs back to $D$.
A shared local projection head
$h_\phi:\mathbb{R}^D\to\mathbb{R}^{D_{\mathrm{local}}}$
maps the predictor outputs to a local contrastive embedding
space of dimension $D_{\mathrm{local}}$, yielding $\ell_2$-normalized patch
predictions $\mathbf{p}_{i,t,f}^v\in\mathbb{R}^{D_{\mathrm{local}}}$. The predictor retains both grid axes, and only predictions at masked positions contribute to the local loss as detailed below.

\Needspace{21\baselineskip}
\subsection{Masked-Patch Prediction \& Supervision}
\label{sec:local-objective}

\begin{wrapfigure}{r}{0.30\textwidth}
    \centering
    \includegraphics[width=\linewidth]{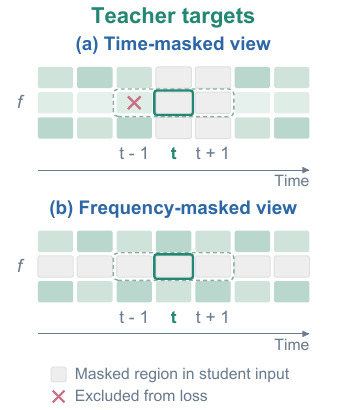}
    \captionsetup{font=small,justification=raggedright}
    \caption{\textbf{Mask-aware target selection.} Temporal masking can exclude
    neighbors; frequency masking retains all in-bounds temporal neighbors.}
    \label{fig:mask-gating}
\end{wrapfigure}

Our local objective uses contrastive self-supervision for masked predictions~\citep{baevski2020wav2vec,yi2023conmim}, with full-view teacher targets and cross-recording negatives. We refer to this local objective as masked latent discrimination (MLD).
We first define prediction-target discrimination at the masked patch's own position, then extend it to eligible temporal neighboring positions. 

\paragraph{Full-view latent targets.}
The teacher encoder $f_{\bar\theta}$ and local projection head $h_{\bar\phi}$ are updated as exponential moving averages (EMA) of the student encoder $f_\theta$ and local projection head $h_\phi$, respectively.
For each student view, the teacher processes the corresponding full spectrogram $X_i^v$ without structured masking (Figure~\ref{fig:framework}).
Its patch representations are projected into
$\ell_2$-normalized targets
$\mathbf{q}_{i,t,f}^v\in\mathbb{R}^{D_{\mathrm{local}}}$,
with gradients stopped through the teacher branch.

\paragraph{Discriminating the corresponding teacher target.}
For a masked anchor $(t,f)\in\mathcal{M}_i^v$, same-position cross-recording contrast is defined as
\begingroup
\small
\begin{equation}
    \ell_i^v(\mathbf{p}_{i,t,f}^v;(t,f))
    = -\log
    \frac{\exp\!\left((\mathbf{p}_{i,t,f}^v)^{\top}\mathbf{q}_{i,t,f}^v/\tau_p\right)}
    {\displaystyle\sum_{j=1}^{B}
        \exp\!\left((\mathbf{p}_{i,t,f}^v)^{\top}\mathbf{q}_{j,t,f}^v/\tau_p\right)},
    \label{eq:position-infonce}
\end{equation}
\endgroup
where $\tau_p>0$ is the local temperature. The positive is the teacher target $\mathbf{q}_{i,t,f}^v$ from recording $i$, and negatives come from other recordings $j\ne i$ at the same grid coordinate $(t,f)$. Only targets at $(t,f)$ enter the denominator; other positions within a recording are not negative candidates. A shared coordinate across recordings does not imply semantic alignment of their sound events.

\paragraph{Temporal neighborhood extension.}
We extend MLD with weaker supervision from nearby teacher targets, motivated by temporal-neighborhood learning \citep{tonekaboni2021unsupervised} and neighborhood distribution smoothing \citep{yang2021delving}. These works motivate a local-consistency prior; our prediction-target discrimination retains the cross-recording InfoNCE formulation above. We define neighborhoods over temporal positions within a recording and keep the frequency coordinate fixed. For an integer radius $K\geq 0$, let $\mathcal{T}=\{-K,\ldots,0,\ldots,K\}$, with offsets measured in temporal patch steps. The corresponding temporal window contains $W=2K+1$ positions. At offset $\delta$, the same prediction $\mathbf{p}_{i,t,f}^v$ is contrasted with targets at $(t+\delta,f)$ through $\ell_i^v(\mathbf{p}_{i,t,f}^v;(t+\delta,f))$. 
Nearby temporal regions may share acoustic content, making teacher targets at neighboring positions within the same recording useful candidates for additional
positive supervision.
We use prior weights $w_\delta$, assigning a positive weight to the reference target and smaller nonnegative weights to neighboring targets, preserving the reference position as the primary prediction target.

\paragraph{Mask-aware neighbor selection.}
If a neighboring position $(t+\delta,f)$ is visible to the student, the predictor can trivially copy its observed features rather than inferring the masked anchor. To prevent such shortcuts, 
as illustrated in Figure~\ref{fig:mask-gating}, a neighboring teacher target contributes to the loss only if its position lies within the grid and is also masked in the anchor recording's student view. For each masked anchor, the valid offsets are
\begin{equation}
    \mathcal{V}_{i,t,f}^v
    = \{\delta\in\mathcal{T}:(t+\delta,f)\in\mathcal{M}_i^v\}.
    \label{eq:mask-gating}
\end{equation}
The center position is always included because the anchor itself is masked. For each retained offset $\delta$, teacher targets $\mathbf{q}_{j,t+\delta,f}^v$ from other recordings $j\ne i$ at the same coordinate serve as negatives.

\paragraph{Weighted loss aggregation.}
We define each anchor's neighborhood loss as a weighted average of the valid offset losses:
\begin{equation}
    \ell_{i,t,f}^v
    = \sum_{\delta\in\mathcal{V}_{i,t,f}^v}
        \tilde{w}_{i,t,f,\delta}^v\,
        \ell_i^v(\mathbf{p}_{i,t,f}^v;(t+\delta,f)),
    \label{eq:weight-renorm}
\end{equation}
where $\tilde{w}_{i,t,f,\delta}^v$ denotes the normalized weight for each valid offset. After excluding invalid offsets, we rescale the remaining prior weights $w_\delta$ to sum to one
for each anchor, preserving their relative proportions.
For each valid offset $\delta$, the contrastive denominator
is computed over recordings at the target coordinate
$(t+\delta,f)$.
When $K=0$, or when no off-center neighbor is valid,
the neighborhood loss reduces to the same-position
contrastive loss in Eq.~\ref{eq:position-infonce}.
The local objective $\mathcal{L}_{\mathrm{MLD}}$ first
averages the anchor losses over all masked patches across
recordings within each view, then averages the two
view-level losses with equal weight.

\subsection{Joint Training Objective}
\label{sec:joint-objective}

We use a symmetric NT-Xent objective for global contrast $\mathcal{L}_{\mathrm{global}}$~\citep{huang2026audiomosaic,chen2020simclr}.
For anchor $\mathbf{z}_i^v$ (e.g., a time-masked view), the positive is the other masked view of the same recording, $\mathbf{z}_i^{3-v}$ (e.g., its corresponding frequency-masked view); both views of other recordings serve as negatives. Each denominator includes the positive and excludes only the anchor itself.
We jointly optimize the student encoder $f_\theta$, predictor $d_\eta$, global projection head, and local projection head $h_\phi$ with
\begin{equation}
    \mathcal{L}
    = \mathcal{L}_{\mathrm{global}}
      +\lambda_{\mathrm{local}}\mathcal{L}_{\mathrm{MLD}},
    \label{eq:total-loss}
\end{equation}
where $\lambda_{\mathrm{local}}$ weights the local loss relative to the global loss. The EMA momentum schedule is specified in Appendix~\S\ref{sec:app-hyperparameters}.
Only the pretrained student encoder is transferred to downstream tasks. The local predictor, both pretraining projection heads, and the teacher are discarded. Task-specific classifiers are trained during downstream probing or fine-tuning.

\section{Experiments}
\label{sec:experiments}
\definecolor{soundfmshade}{RGB}{235,243,248}

\subsection{Experimental Setup}
\label{sec:exp-setup}

\paragraph{Pretraining.}
We pretrain our model on the widely used AudioSet corpus~\citep{gemmeke2017audio} without labels. Due to YouTube video deletions over time, our downloaded version consists of approximately 1.68M valid audio clips ($\sim 4{,}665$ hours), representing a \textbf{smaller pretraining corpus} than the 1.91M-2.0M clips typically employed in prior SSL literature~\citep{huang2022audiomae,chen2024eat,huang2026audiomosaic,alex2025sslam,chen2023beats}. 

\paragraph{Evaluation protocols.}
We assess downstream performance and transferability under three complementary settings.
\textbf{Full fine-tuning:}
We jointly fine-tune the encoder and task-specific classification heads on standard benchmarks: AudioSet-2M, AudioSet-20K, 
ESC-50, Speech Commands V1/V2, and FSD50K.
Subject to the same attrition, our retained AudioSet evaluation split comprises 16,103 clips.
\textbf{Linear probing and domain transfer:}
We evaluate frozen representations across general audio, bioacoustics, medical acoustics, and speech emotion recognition.
For baseline encoders, we obtain open-weight self-supervised pretrained checkpoints and evaluate them alongside \modelname{} using a shared evaluation pipeline.
Within each task, we use consistent data splits, label definitions, evaluation metrics, and probe-training protocols across encoders.
Encoder parameters remain frozen, and only the downstream probe is optimized.
We consider two configurations: a linear classifier applied directly to frozen features (Bare), and a linear classifier preceded by LayerNorm (LN).
Each model-dataset-probe configuration is evaluated over three random seeds, and we report the mean performance across runs.
Dataset protocols and optimization settings are detailed in Appendix~\S\ref{sec:app-linear-probing} and Appendix~\S\ref{sec:app-hyperparameters}.
\textbf{Zero-shot transfer:}
We evaluate zero-shot transfer through sound event tagging and emotion recognition, where shared label spaces allow source-trained predictors to be applied across datasets
without target-domain training.
We also examine medical-audio embeddings to explore whether general audio pretraining qualitatively captures structure relevant to physiological sounds.

\paragraph{Baselines.}
We compare with recent open-source audio representation learning methods,
including AudioMAE~\citep{huang2022audiomae},
BEATs~\citep{chen2023beats}, EAT~\citep{chen2024eat},
SSLAM~\citep{alex2025sslam}, BAT~\citep{ghaffari2026bat},
and AudioMosaic~\citep{huang2026audiomosaic}.
For full fine-tuning comparisons, we re-evaluate available author-released fine-tuned
checkpoints on our retained AS-2M evaluation subset without further training. For AS-20K and FSD50K, we fine-tune released pretrained checkpoints on our training splits and evaluate them on the same evaluation splits as our model. Since BAT reports no SPC-1 results and uses a 12-class SPC-2 setup rather than the 35-class setup used by the other baselines, we additionally fine-tune it on both tasks.
Other baseline fine-tuning results are taken from the original papers
unless otherwise indicated.

\paragraph{Model and training configuration.}
Unless otherwise specified, \modelname{} uses ViT-Base encoders~\citep{dosovitskiy2020image} for both the student and EMA teacher.
Audio is resampled to 16~kHz mono and converted into 128-band log-Mel spectrograms with 1,024 time frames.
We divide each spectrogram into non-overlapping $16\times16$ patches and construct two complementary views by masking 60\% of temporal slices in one view and 40\% of frequency-patch bands in the other~\citep{huang2026audiomosaic}. The default temporal-neighborhood radius is $K=1$.
Student parameters are optimized with AdamW~\citep{loshchilov2017decoupled} and a cosine-annealed learning rate scheduler. We pretrain our models on a Cloud TPU v4-128 slice (64 chips) with a batch size of 6144. Detailed architecture specifications, training hyperparameters and downstream evaluation settings are provided in Appendix~\S\ref{sec:app-hyperparameters}.

\subsection{Main Results}
\label{sec:exp-main}

\paragraph{Supervised fine-tuning.}
We first assess \modelname{}’s adaptability through supervised fine-tuning on standard audio and speech benchmarks (Table~\ref{tab:literature-finetune-audiomosaic}). \modelname{} achieves leading results among the compared methods across sound classification and keyword spotting. Its improvements over AudioMosaic extend to both AudioSet-20K and FSD50K, with a gain of 0.9 mAP on the latter. On AudioSet-2M, \modelname{} matches the highest published mAP (50.2) and outperforms all re-evaluated checkpoints on the shared evaluation subset (e.g., +1.0 mAP over AudioMosaic), while using approximately 13\% fewer pretraining and fine-tuning clips.

\begin{table}[t]
\centering
\caption{\textbf{Comparison with representative audio self-supervised learning methods.}
AS-20K, AS-2M, and FSD50K report mAP; the other tasks report accuracy (\%).
AS-2M baseline entries show published / re-evaluated mAP.
Re-evaluation uses released fine-tuned checkpoints on the same evaluation subset as our model, without additional training.
PT/FT: pre-training/fine-tuning; AS: AudioSet.
AS$^{*}$: our smaller available AudioSet subset ($\sim$1.68M clips).
$^{\ddagger}$: Re-fine-tuned on our AudioSet-20K training split and evaluated on our retained evaluation subset.}
\label{tab:literature-finetune-audiomosaic}
\begingroup
\footnotesize
\setlength{\tabcolsep}{2.2pt}
\renewcommand{\arraystretch}{1.08}
\resizebox{0.94\textwidth}{!}{%
\begin{tabular}{@{}lcccccccccc@{}}
\toprule
\multirow{2}{*}[-0.65ex]{\textbf{Model}} & \multirow{2}{*}[-0.65ex]{\textbf{Backbone}} &
\multirow{2}{*}[-0.65ex]{\shortstack{\textbf{Data}\\(PT)}} &
\multicolumn{2}{c}{\textbf{Params}} &
\multicolumn{4}{c}{\textbf{Audio}} & \multicolumn{2}{c}{\textbf{Speech}} \\
\cmidrule(lr){4-5}\cmidrule(lr){6-9}\cmidrule(l){10-11}
 & & & PT & FT & AS-20K & AS-2M & FSD50K & ESC-50 & SPC-2 & SPC-1 \\
\midrule
BEATs$_{iter3}$~\citep{chen2023beats} & ViT-B/16 & AS & 182M & 91M & 38.0$^{\ddagger}$ & 48.0 / 46.7 & 57.6 & 95.6 & 98.3 & 97.7 \\
AudioMAE~\citep{huang2022audiomae} & ViT-B/16 & AS & 137M & 86M & 35.5$^{\ddagger}$ & 47.3 / 46.5 & 57.3 & 94.1 & 98.3 & 96.9 \\
A-JEPA~\citep{fei2023jepa} & ViT-B/16 & AS & 354M & 86M & 38.4 & 48.6 / - & - & 96.3 & \textbf{98.5} & 97.7 \\
EAT~\citep{chen2024eat} & ViT-B/16 & AS & 93M & 88M & 40.5$^{\ddagger}$ & 48.6 / 48.8 & 67.5 & 95.9 & 98.3 & - \\
SSLAM~\citep{alex2025sslam} & ViT-B/16 & AS & 93M & 88M & 39.3$^{\ddagger}$ & \textbf{50.2} / 50.0 & 66.6 & 96.2 & 98.1 & 98.8 \\
BAT~\citep{ghaffari2026bat} & ViT-B/16 & AS & 93M & 88M & 40.3$^{\ddagger}$ & 48.9 /  -  & 68.0 & 96.0 & 98.2 & 98.8 \\
AudioMosaic~\citep{huang2026audiomosaic} & ViT-B/16 & AS & 86M & 86M & 41.8$^{\ddagger}$ & \textbf{50.2} / 49.2 & 70.5 & 97.5 & 98.4 & 99.0 \\
\midrule
\rowcolor{soundfmshade}
\textbf{\modelname} & ViT-B/16 & AS$^{*}$ & 94M & 86M & \textbf{42.3} & \textbf{50.2} & \textbf{71.4} & \textbf{98.2} & \textbf{98.5} & \textbf{99.1} \\
\bottomrule
\end{tabular}%
}
\endgroup
\end{table}

\paragraph{Linear probing and domain transfer.}

\begin{table*}[t]
\centering
\scriptsize
\setlength{\tabcolsep}{1.8pt}
\caption{\textbf{Linear Probing Performance}. We compare frozen representations evaluated under \textbf{Strict Bare Probing} (raw features without normalization) versus \textbf{Flexible Probing} (with downstream LayerNorm). Results are means over three random seeds, and all metrics are percentages (\%). Best score is in \textbf{bold}, and the runner-up is \underline{underlined}.}
\label{tab:cross-domain-probing}
\resizebox{\textwidth}{!}{%
\begin{tabular}{llc ccccc >{\columncolor{soundfmshade}}c ccccc >{\columncolor{soundfmshade}}c}
\toprule
& & & \multicolumn{6}{c}{\textbf{Strict Bare Probing (Raw Features)}} & \multicolumn{6}{c}{\textbf{Flexible Probing (+LayerNorm)}} \\
\cmidrule(lr){4-9}\cmidrule(lr){10-15}
\textbf{Domain} & \textbf{Dataset} & \textbf{Metric} & A-MAE & EAT & SSLAM & BAT & AMos & \textbf{\modelname} & A-MAE & EAT & SSLAM & BAT & AMos & \textbf{\modelname} \\
\midrule
\multirow{3}{*}{\textbf{General}} 
& AudioSet-20K         & mAP      & 13.9 &  1.1 &  0.9 & 18.1 & \underline{28.7} & \textbf{31.2} & 11.9 & 18.7 & 18.2 & 28.5 & \underline{33.4} & \textbf{35.0} \\
& FSD50K               & mAP      & 32.7 & 31.4 & 30.3 & \underline{53.9} & 53.7 & \textbf{55.3} & 33.8 & 39.3 & 37.0 & 53.3 & \underline{55.4} & \textbf{58.6} \\
& ESC-50               & Acc      & 77.5 & 66.6 & 87.0 & 92.7 & \underline{93.2} & \textbf{94.5} & 77.0 & 83.4 & 91.5 & 92.4 & \underline{93.8} & \textbf{94.7} \\
\midrule
\multirow{3}{*}{\textbf{Bioacoustics}} 
& BEANS-Watkins        & Acc      & 81.9 & 61.9 & 34.5 & 57.1 & \underline{88.9} & \textbf{89.1} & 70.7 & 85.6 & 82.3 & 82.2 & \textbf{89.3} & \underline{85.9} \\
& BEANS-Dogs           & Acc      & \underline{86.1} & 60.9 & 37.4 & 59.5 & \textbf{88.7} & 85.9 & 74.6 & \textbf{85.4} & 78.7 & 82.5 & \underline{85.1} & 83.5 \\
& BirdSet-HSN          & mAP      & \underline{12.4} &  7.0 &  6.1 &  5.0 & \textbf{13.1} & 11.7 &  9.6 & 11.1 &  9.9 &  9.4 & \underline{11.8} & \textbf{11.9} \\
\midrule
\multirow{3}{*}{\textbf{Medical}} 
& PSG-Audio (Binary)   & Bal.\ Acc & 71.8 & 70.2 & 67.6 & 75.1 & \underline{77.2} & \textbf{79.1} & 69.7 & 75.6 & 74.2 & \underline{77.5} & 75.6 & \textbf{78.1} \\
& PSG-Audio (Subtype)  & Macro-F1 & 30.4 & 21.3 & 22.9 & 26.7 & \underline{34.7} & \textbf{37.9} & 26.4 & 31.7 & 32.7 & \textbf{35.5} & 31.4 & \underline{33.1} \\
& CirCor Murmur        & Macro-F1 & \underline{68.0} & 28.4 & 28.4 & 52.2 & 66.7 & \textbf{72.7} & \underline{68.1} & 62.4 & 59.7 & 65.8 & 62.0 & \textbf{68.9} \\
\midrule
\multirow{3}{*}{\textbf{Speech Emotion}} 
& CREMA-D              & Macro-F1 & 64.6 & 51.1 & 39.3 & 52.1 & \underline{68.2} & \textbf{68.7} & 56.7 & \underline{65.4} & 60.6 & \textbf{69.8} & 63.6 & 64.9 \\
& RAVDESS              & Macro-F1 & \underline{57.5} & 41.5 & 34.9 & 40.5 & 53.5 & \textbf{58.9} & 47.0 & 54.0 & 49.4 & \textbf{62.2} & 53.8 & \underline{56.8} \\
& TESS                 & Macro-F1      & \underline{99.0} & 86.4 & 78.6 & 84.1 & \textbf{99.8} & \textbf{99.8} & 91.2 & \underline{99.2} & 98.3 & 99.1 & \textbf{99.5} & \textbf{99.5} \\
\bottomrule
\end{tabular}%
}
\end{table*}

Beyond adaptation through fine-tuning, a general-purpose audio encoder can also serve as an off-the-shelf feature extractor for diverse downstream tasks. By keeping the encoder fixed and standardizing downstream training, linear probing helps isolate the contribution of pretrained representations from differences in fine-tuning recipes, including augmentation and optimization choices. We evaluate \modelname{} and baseline encoders across diverse acoustic domains using a shared linear-probing protocol, assessing how readily task-relevant information can be extracted from their frozen features. As shown in Table~\ref{tab:cross-domain-probing}, \modelname{} matches or outperforms the evaluated baselines on the majority of tasks under both probing protocols. Its advantage is particularly evident under bare probing, where it achieves the best or joint-best performance on 10 of 12 tasks without downstream feature normalization. \modelname{} leads all three general-audio benchmarks under both protocols, while its bare features also achieve the highest scores on all three medical-acoustic tasks. On PSG-Audio subtype classification and CirCor murmur classification, it improves macro-F1 over the strongest baseline by 3.2 and 4.7 percentage points, respectively. These superior results suggest that \modelname{} is able to learn rich semantic audio representations.

\begin{figure}[!htbp]
\begin{minipage}[t]{0.46\linewidth}
\vspace{0pt}
\paragraph{Fine-grained sound event detection.}
\label{sec:exp-temporal}
\mbox{}\par\noindent
Beyond clip-level recognition, we evaluate whether the frozen representations retain the temporal detail needed to detect sound events at finer temporal resolution on AudioSet-Strong. \modelname{} outperforms the evaluated baselines across frame mAP, PSDS1, and PSDS2 (Figure~\ref{fig:sed-audioset-strong}). These gains extend the benefits observed in clip-level classification to both frame-level prediction and event-based detection. Dataset and metric details are provided in Appendix~\S\ref{sec:app-frame-probing}.
\end{minipage}\hfill
\begin{minipage}[t]{0.50\linewidth}
\vspace{0pt}
\centering
\includegraphics[width=\linewidth]{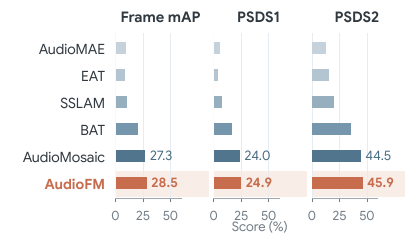}
\caption{\textbf{Frozen-encoder SED on AS-Strong.} }
\label{fig:sed-audioset-strong}
\end{minipage}
\end{figure}

\begin{figure}[!htbp]
\centering
\includegraphics[width=\linewidth]{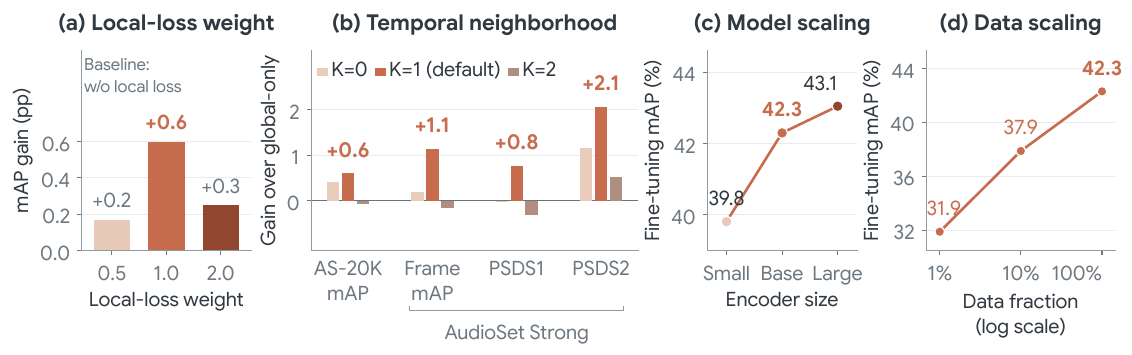}
\caption{\textbf{Ablations and scaling.}
(a) Local-loss weight $\lambda_{\mathrm{local}}$ with the default temporal neighborhood ($K=1$).
(b) Temporal-neighborhood radius.
Panels (a,b) report frozen-encoder probing gains over training without local loss;
(c,d) AS-20K fine-tuning performance as encoder size and pretraining data increase.}
\label{fig:ablations-scaling}
\end{figure}

\FloatBarrier
\begin{samepage}
\subsection{Ablation Studies and Analysis}
\label{sec:exp-ablation}
We examine local supervision through frozen-encoder probing, and assess model and data scaling through AS-20K fine-tuning (Figure~\ref{fig:ablations-scaling}). Full neighborhood configurations and masking-ratio sensitivity results are provided in Appendix~\S\ref{sec:app-hyperparameters} and Appendix~\S\ref{sec:app-masking-sensitivity}, respectively.

\end{samepage}

\paragraph{Model design.}
Adding masked-patch supervision improves AS-20K probing over global-only training
across all tested positive local-loss weights, with
$\lambda_{\mathrm{local}}=1.0$ performing best
(Figure~\ref{fig:ablations-scaling}(a)).
Within the local objective, the compact neighborhood ($K=1$) outperforms
center-only supervision on both AS-20K classification and AudioSet Strong
detection, including frame mAP and both PSDS metrics
(Figure~\ref{fig:ablations-scaling}(b)).
This supports using immediate temporal neighbors as additional local supervision.
Extending the window further to $K=2$, which distributes more weight to distant offsets, reduces performance, indicating that neighborhood supervision benefits primarily from immediate temporal neighbors.

\paragraph{Scaling with model size and data.}
We observe consistent scaling gains in AS-20K fine-tuning with increasing
encoder capacity and pretraining data
(Figure~\ref{fig:ablations-scaling}(c,d)).
Performance increases from Small to Base to Large, while increasing the pretraining data (1\% to 100\%) yields consistent gains across the two orders of magnitude evaluated, supporting joint global-local masked discriminative pretraining as a scalable audio representation learning objective.

\FloatBarrier
\subsection{Cross-Domain Generalization}
\label{sec:exp-zeroshot}
\begin{figure}[!htbp]
\centering
\includegraphics[width=\linewidth]{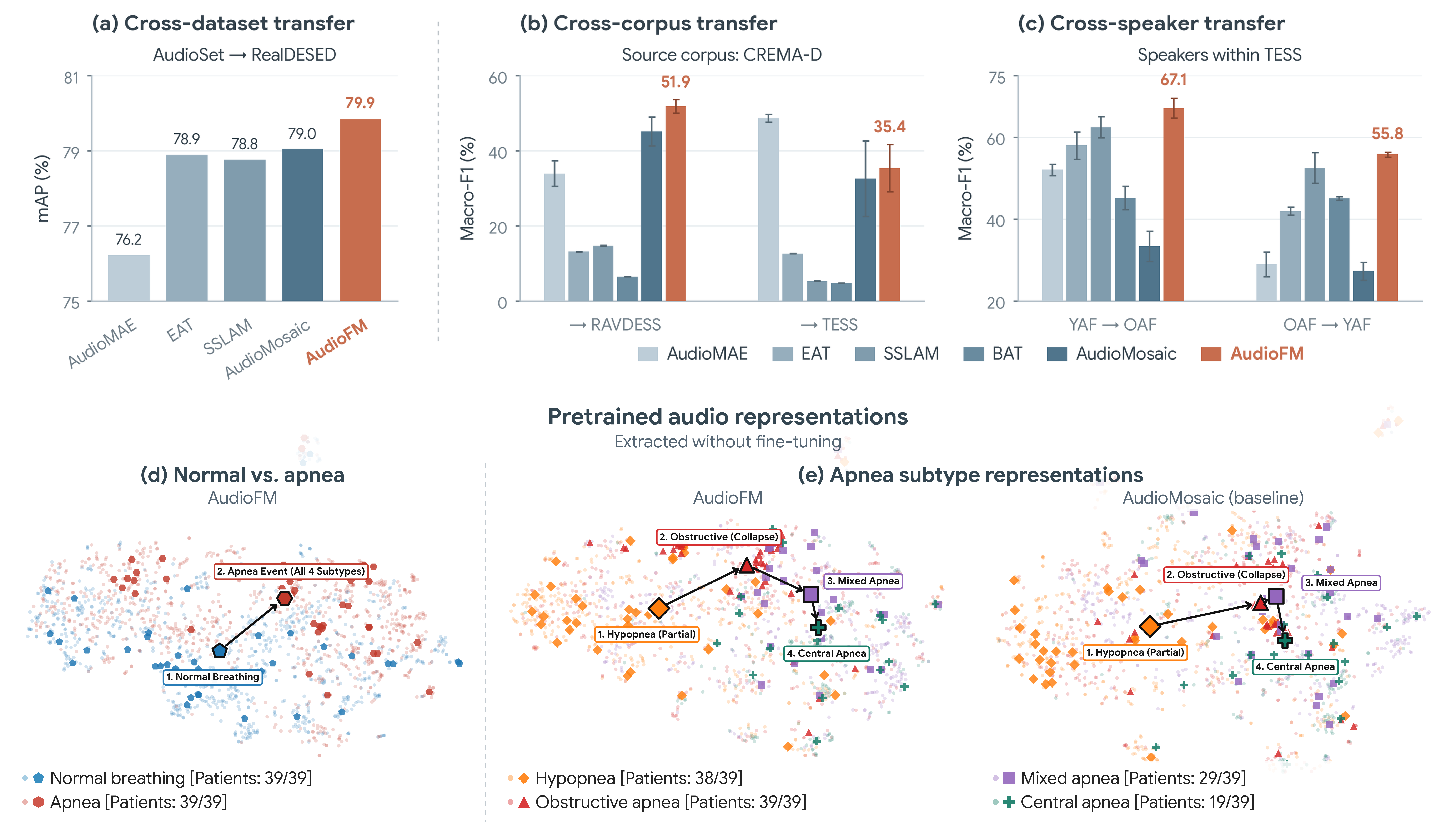}
\caption{\textbf{Cross-domain transfer and pretrained audio representations.}
(a-c) Direct transfer without target-domain training: (a) clip-level sound event tagging on RealDESED, (b) cross-corpus and (c) cross-speaker emotion recognition. Error bars indicate standard deviations over three seeds.
(d-e) t-SNE visualizations of PSG-Audio embeddings extracted from pretrained encoders without fine-tuning: (d) normal breathing versus apnea using \modelname{}, and (e) four apnea subtypes using \modelname{} and AudioMosaic.}
\label{fig:cross-domain-transfer}
\end{figure}

We evaluate whether source-trained predictors transfer to new datasets and speakers without target-domain training, and further examine medical-audio embeddings extracted directly from pretrained encoders to provide a qualitative view of their representations in a specialized acoustic domain (Figure~\ref{fig:cross-domain-transfer}).

\paragraph{Sound event predictors transfer across datasets.}
Here we apply AudioSet-trained classifiers directly to RealDESED~\citep{schmid2026realdesed}, mapping their outputs to 12 shared event categories while keeping both the encoder and classification head fixed.
\modelname{} achieves 79.9\% mAP, outperforming all evaluated baselines (Figure~\ref{fig:cross-domain-transfer}(a)).
This result demonstrates that the source-trained predictor remains effective on domestic soundscapes without target-specific adaptation. The class mapping is provided in Appendix~\S\ref{sec:app-desed-breakdown}.

\paragraph{Emotion recognition across corpora and speakers.}
We next examine whether emotion classifiers trained on one corpus
or speaker remain effective on another.
We train linear emotion classifiers on frozen source-domain representations and evaluate them directly on unseen corpora or speakers.
The evaluation covers CREMA-D $\to$ RAVDESS / TESS and bidirectional transfer between the two TESS speakers.
\modelname{} ranks first on three of the four transfer tasks and second on the remaining task, with particularly strong results on CREMA-D $\to$ RAVDESS and both cross-speaker directions (Figure~\ref{fig:cross-domain-transfer}(b,c)).
These results indicate that its representations capture emotion-relevant information that transfers beyond the source corpus and speaker.
Detailed protocols are provided in Appendix~\S\ref{sec:app-zeroshot-emotion}.

\paragraph{Frozen representation visualization reflects domain-relevant structure.}
Figure~\ref{fig:cross-domain-transfer}(d-e) examines how \modelname{}'s pretraining shapes the zero-shot representations of out-of-domain medical acoustics (PSG-Audio test cohort). \modelname{} distinguishes normal breathing from apnea. Notably, when zooming into the four apnea subtypes, \modelname{} exhibits a clearer arrangement of subtype centers in the projected embedding space compared with the baseline, spanning hypopnea, obstructive, mixed, and central apnea, motivating further exploration of its use in medical acoustic analysis.

\FloatBarrier
\section{Discussion}
\label{sec:conclusion}

\textbf{Limitations.}
Our pretraining and scaling analyses focus on AudioSet and a limited range of model capacities, with all recordings converted to single-channel audio.
Extending this framework to larger, more diverse corpora and multichannel inputs that preserve spatial cues remains an open direction.
While our evaluation spans four acoustic domains and includes cross-corpus transfer analysis across different acoustic environments, robustness to 
extreme channel distortion and controlled device shifts remains underexplored.
Future work could investigate transfer under a broader range of recording conditions and extend the evaluation
to additional specialized tasks. In particular, our medical-audio findings highlight the potential of pretrained audio representations for physiological sound analysis, motivating future studies across broader tasks and clinical settings.

\textbf{Conclusion.}
We introduced \modelname{}, a self-supervised audio framework that combines recording-level cross-view contrast with masked latent discrimination on a two-dimensional time-frequency grid. Predicted masked-patch representations identify their corresponding full-view teacher targets among candidates from other recordings. The temporal-neighborhood extension retains this reference-position supervision while incorporating weaker supervision from eligible neighbors. \modelname{} delivers leading performance across a broad range of audio tasks, matching or outperforming the evaluated state-of-the-art audio SSL baselines on the majority of frozen-encoder classification tasks and surpassing them across all reported temporal localization metrics. Beyond standard fine-tuning benchmarks, our study provides a systematic cross-domain evaluation through 12 frozen-encoder classification tasks spanning general audio, bioacoustics, medical acoustics, and speech emotion, complemented by frame-level sound event detection and cross-corpus and cross-speaker emotion transfer. This combined coverage examines representation quality across acoustic domains, temporal resolutions, and adaptation settings. Together with strong fine-tuning and cross-domain transfer results, these findings demonstrate the effectiveness of combining recording-level alignment with discriminative prediction of masked local representations.

\bibliography{ref}
\bibliographystyle{plainnat}

\appendix
\clearpage
\section{Appendix}
\label{sec:appendix}

\subsection{Model and Training Details}
\label{sec:app-hyperparameters}

Table~\ref{tab:app-training-hyperparameters} summarizes optimization settings by dataset. Input, model, and pretraining-objective configurations follow the table.

\begin{table}[!htbp]
\centering
\fontsize{8}{9.5}\selectfont
\caption{\textbf{Pre-training (PT) and fine-tuning (FT) hyperparameters} for \modelname{}.}
\label{tab:app-training-hyperparameters}
\begingroup
\setlength{\tabcolsep}{3pt}
\renewcommand{\arraystretch}{1.05}
\begin{tabular}{@{}l*{7}{c}@{}}
        \toprule
        & Pre-training & \multicolumn{6}{c}{Fine-tuning} \\
        \cmidrule(lr){2-2}\cmidrule(lr){3-8}
        Configuration & AS-2M PT  & AS-2M  & AS-20K  & FSD50K & ESC & SPC-2  & SPC-1 \\
        \midrule
        Optimizer & \multicolumn{7}{c}{AdamW}\\
        Optimizer momentum & \multicolumn{7}{c}{$\beta_1=0.9$, $\beta_2=0.999$}\\
        Weight decay & 0.05 & \multicolumn{6}{c}{0.01} \\
        LR layer decay & 1.0 & 0.75 & 0.75 & 0.75 & 0.75 & 0.75 & 0.75 \\
        Peak learning rate & 6e-4 & 1.5e-4 & 5e-5 & 5e-5 & 5e-5 & 5e-4 & 5e-4 \\
        Learning rate schedule & \multicolumn{7}{c}{Half-cycle cosine decay}\\
        Minimum learning rate & 1e-6 & 3e-7 & 1e-7 & 1e-7 & 1e-7 & 1e-6 & 1e-6 \\
        Gradient clipping & 1.0 & \multicolumn{6}{c}{None}\\
        Warm-up steps & 2,500 & 7,812 & 2,000 & 2,000 & 2,000 & 3,300 & 2,000 \\
        Training steps & 125,000 & 78,125 & 100,000 & 200,000 & 50,000 & 132,600 & 40,000 \\
        Batch size & 6144 & 256 & 64 & 128 & 32 & 128 & 128 \\
        Weighted sampling  & False & True & False & False & False & False & False \\
        Weighted sampling size & - & 200,000 & - & - & -& - & - \\
        Augmentation & R & R & R & R & R & R+N & R+N \\
        SpecAug (time/frequency) & - & 192/48 & 192/48 & 192/48 &  96/24 & 48/48 & 48/48 \\
        Drop path & 0.1 & 0.1 & 0.1 & 0.1 & 0.1 & 0.1 & 0.1 \\
        Dropout & 0.0 & 0.0 & 0.0 & 0.0 & 0.0 & 0.0 & 0.0 \\
        Mixup & 0.0 & 0.8 & 0.8 & 0.8 & 0.0 & 0.8 & 0.8 \\
        Multilabel & N/A & True& True& True & False & False & False \\
        Loss Function & NT-Xent + MLD & BCE & BCE & BCE & CE & BCE & BCE \\
        Normalization mean & -4.268 & -4.268 & -4.268 & -6.846 & -6.627 & -6.846 & -6.846 \\
        Normalization std. & 4.569 & 4.569 & 4.569 & 5.565 & 5.359 & 5.565 & 5.565 \\ \bottomrule
\end{tabular}
\endgroup
\par\smallskip
\begin{minipage}{\linewidth}
\fontsize{8}{9.5}\selectfont
R: random temporal start with cyclic rolling; N: spectrogram noise at 20\,dB SNR.
BCE: binary cross entropy; CE: cross entropy; NT-Xent: normalized temperature-scaled cross entropy; MLD: masked latent discrimination.

\end{minipage}
\end{table}

\paragraph{Model architecture.}
The encoder is a ViT-Base with 12 layers, hidden dimension $D=768$, 12 attention heads, and learnable 2D positional embeddings. The shared predictor has four Pre-LN Transformer blocks at bottleneck dimension $D_{\mathrm{pred}}=384$ with six attention heads, input/output projections between $D$ and $D_{\mathrm{pred}}$, a final LayerNorm, and masking-branch embeddings. Its axis-separated 2D SinCos position encoding concatenates 192-dimensional temporal and frequency components. The global projector outputs 128-dimensional normalized embeddings. The student and EMA teacher local projection heads share the same architecture: a two-layer MLP with LayerNorm, dimensions $768\to512\to128$, and output $\ell_2$ normalization ($D_{\mathrm{local}}=128$).

\paragraph{Pretraining objective and updates.}
We combine $\mathcal{L}_{\mathrm{global}}$ and $\mathcal{L}_{\mathrm{MLD}}$ with $\lambda_{\mathrm{local}}=1.0$, using temperatures $\tau_g=0.1$ and $\tau_p=0.2$. The reference neighborhood radius is $K=1$ (window width $W=3$), with prior weights $(w_{-1},w_0,w_{+1})=(0.1,0.8,0.1)$. Each temporal patch step corresponds to 160~ms, so side targets are at offsets $\pm160$~ms with frequency held fixed. Weights are renormalized over mask-aware valid offsets using Eq.~\ref{eq:weight-renorm}: if only one side neighbor is valid, the center and neighbor receive weights $8/9$ and $1/9$; if neither is valid, the center receives weight $1$. Teacher parameters use EMA momentum increasing from $0.990$ to $0.9995$. Student parameters use AdamW with linear warm-up followed by cosine decay. The remaining optimization settings are given in Table~\ref{tab:app-training-hyperparameters}.

\paragraph{Neighborhood ablation configurations.}
The center-only MLD setting corresponds to $K=0$ ($W=1$). The default $K=1$ setting uses prior weights $(0.1,0.8,0.1)$, equivalent to a Gaussian sampled at offsets $\{-1,0,1\}$ and normalized on this support. The wider $K=2$ ($W=5$) variant uses prior weights $(0.03,0.20,0.54,0.20,0.03)$ at temporal offsets $(-2,-1,0,1,2)$, keeping frequency fixed.

\paragraph{Linear probing.}
For the utterance-level probing benchmark (Table~\ref{tab:cross-domain-probing}), we evaluate both a strict linear classifier on frozen encoder features (Bare) and a linear classifier preceded by LayerNorm (Flexible). All probes are trained with AdamW ($\beta_1 = 0.95$, $\beta_2 = 0.999$, weight decay $5 \times 10^{-7}$) using a peak learning rate of $1 \times 10^{-3}$ with linear warm-up followed by decay to $1 \times 10^{-5}$. Training steps and batch sizes are scaled to corpus size: 200,000 steps (batch 128) for FSD50K; 50,000 steps (batch 32) for ESC-50; 15,000 steps (batch 64) for BirdSet-HSN; 10,000 steps for AudioSet-20K (batch 128) and PSG-Audio Binary/Subtype (batch 128/64); 5,000 steps for BEANS-Watkins, BEANS-Dogs, CirCor Murmur (batch 32), and CREMA-D (batch 64); and 3,000 steps (batch 32) for RAVDESS and TESS. Short-clip tasks (ESC-50, BEANS-Watkins, and BirdSet-HSN) use a 5.12-second input window, whereas all remaining tasks use a 10.24-second window. No data augmentations (e.g., Mixup or SpecAugment) are applied during linear probing.

\subsection{Masking-Ratio Sensitivity}
\label{sec:app-masking-sensitivity}

\begin{figure}[!htbp]
\centering
\includegraphics[width=0.70\linewidth]{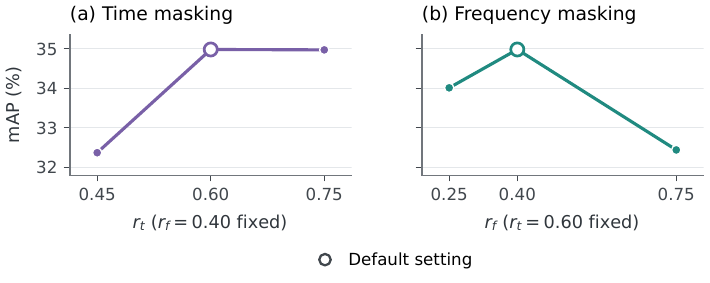}
\caption{\textbf{Sensitivity to structured masking ratios.} AudioSet-20K linear-probing mAP when varying (a) $r_t$ with $r_f=0.40$ fixed, and (b) $r_f$ with $r_t=0.60$ fixed. Both panels use the same vertical scale. Hollow markers identify the shared default configuration $(r_t,r_f)=(0.60,0.40)$.}
\label{fig:masking-ratio-sensitivity}
\end{figure}

We vary the temporal and frequency masking ratios separately around the default configuration $(r_t,r_f)=(0.60,0.40)$ and evaluate the resulting representations using AudioSet-20K linear probing (Figure~\ref{fig:masking-ratio-sensitivity}).
Increasing the temporal masking ratio above the default retains nearly identical performance, whereas reducing it lowers performance.
For frequency masking, both tested alternatives to the default reduce performance, with a larger decrease at the higher ratio.

\subsection{Linear Probing: Datasets and Evaluation Protocols}
\label{sec:app-linear-probing}
\label{sec:app-datasets}

We assess frozen audio representations at two temporal resolutions: utterance-level classification and frame-level sound event detection. Throughout this section, \emph{utterance-level} refers to predictions for an entire audio sample, including both speech and non-speech recordings. The encoder remains frozen while the downstream classifier is trained on labeled examples. Table~\ref{tab:app-dataset-specs} summarizes the task coverage; optimization settings are given in \S\ref{sec:app-hyperparameters}.

\begin{table}[!htbp]
\centering
\small
\caption{\textbf{Linear probing datasets and prediction targets.} Utterance-level tasks span four acoustic domains; AudioSet Strong additionally supports temporal event detection.}
\label{tab:app-dataset-specs}
\begingroup
\setlength{\tabcolsep}{4pt}
\renewcommand{\arraystretch}{1.08}
\begin{tabular}{@{}p{0.19\textwidth}p{0.31\textwidth}p{0.43\textwidth}@{}}
\toprule
\textbf{Domain} & \textbf{Dataset} & \textbf{Prediction target} \\
\midrule
\multicolumn{3}{@{}l}{\textit{Utterance-level classification}} \\
General audio & AudioSet-20K & Sound event categories \\
 & ESC-50 & Environmental sound class \\
 & FSD50K & Sound event categories \\
\midrule
Bioacoustics & BEANS-Watkins & Marine mammal species \\
 & BEANS-Dogs & Individual dog identity \\
 & BirdSet-HSN & Bird species present in a soundscape \\
\midrule
Medical acoustics & PSG-Audio (Binary / Subtype) & Sleep apnea screening and subtype classification \\
 & CirCor Murmur & Patient-level heart murmur classification \\
\midrule
Speech emotion & CREMA-D / RAVDESS / TESS & Expressed emotion \\
\midrule
\multicolumn{3}{@{}l}{\textit{Frame-level sound event detection}} \\
General audio & AudioSet Strong & Event class and temporal extent \\
\bottomrule
\end{tabular}
\endgroup
\end{table}

\subsubsection{Utterance-Level Linear Probing}
\label{sec:app-utterance-probing}

For each utterance-level task, the classifier maps a frozen audio representation to a single class or a set of labels. Probe training settings are summarized in \S\ref{sec:app-hyperparameters}. Dataset descriptions below specify the prediction problem, retained sample counts, and evaluation protocol.

\paragraph{General Audio.}\mbox{}\par
\begin{itemize}[leftmargin=1.5em, labelsep=0.5em, itemsep=0.35em, topsep=0.3em, parsep=0pt, partopsep=0pt, beginpenalty=10000]
\item \textbf{AudioSet-20K.}
AudioSet~\citep{gemmeke2017audio} comprises approximately 10-second YouTube excerpts annotated across 527 sound event classes. For linear probing, we train on the balanced AudioSet-20K subset (16,608 retained clips) and evaluate multi-label recognition on the AudioSet evaluation split (16,103 retained clips) using class-averaged mean average precision (macro mAP).

\item \textbf{ESC-50.}
ESC-50~\citep{piczak2015esc} contains 2,000 five-second environmental recordings organized into 50 balanced classes (40 clips per class). Following the official five-fold partition (1,600 training and 400 test clips per fold), which keeps excerpts from the same source recording within a single fold, we report top-1 classification accuracy averaged across the five folds.

\item \textbf{FSD50K.}
FSD50K~\citep{fonseca2021fsd50k} contains variable-duration Freesound recordings with multi-label annotations across 200 sound event categories. The official development set is subdivided into training and validation portions, alongside a separate evaluation set (10,231 clips). For our FSD50K linear-probing and fine-tuning experiments, we combine the official training and validation portions together (40,965 retained clips) for training with fixed hyperparameters, and evaluate on the official evaluation split (10,231 clips) using macro mAP.
\end{itemize}

\Needspace{4\baselineskip}
\paragraph{Bioacoustics.}\mbox{}\par
We use the BEANS-Watkins, BEANS-Dogs, and BirdSet-HSN benchmarks as integrated in BioFoundation~\citep{hagiwara2023beans,rauch2025birdset,schwinger2025biofoundation}.
\begin{itemize}[leftmargin=1.5em, labelsep=0.5em, itemsep=0.35em, topsep=0.3em, parsep=0pt, partopsep=0pt, beginpenalty=10000]
\item \textbf{BEANS-Watkins.}
Derived from the Watkins Marine Mammal Sound Database, this benchmark defines a 31-class marine mammal species classification task over underwater vocalizations (1,017 training and 339 test clips), evaluated by top-1 accuracy.

\item \textbf{BEANS-Dogs.}
This task contains domestic dog barks annotated by individual identity across ten dogs (415 training and 139 test clips). We treat it as 10-class individual identification and report top-1 accuracy.

\item \textbf{BirdSet-HSN.}
The High Sierra Nevada (HSN) benchmark covers 21 bird species, pairing focal Xeno-Canto training recordings (5,460 five-second segments) with annotated 5-second soundscape evaluation windows (12,000 test segments). Because multiple species can co-occur within a window, we train with binary cross-entropy and report macro mAP.
\end{itemize}

\paragraph{Medical Acoustics.}\mbox{}\par
These tasks evaluate transfer to clinical respiratory and cardiac recordings under strict patient-disjoint partitions.
\begin{itemize}[leftmargin=1.5em, labelsep=0.5em, itemsep=0.35em, topsep=0.3em, parsep=0pt, partopsep=0pt, beginpenalty=10000]
\item \textbf{PSG-Audio (Binary and Subtype).}
PSG-Audio~\citep{korompili2021psgaudio} pairs overnight polysomnography with synchronized tracheal/ambient microphone recordings across 192 patients, segmented into 10-second windows at 16~kHz (103,210 segments total). 
We construct a patient-disjoint split stratified by per-patient apnea burden: 134 training patients (70,825 segments), 19 validation patients (11,036 segments), and 39 test patients (21,349 segments). 
To account for inter-subject recording gain variations while preserving relative amplitude changes between apneic phases and recovery breaths, each segment is normalized by its patient's full-night root-mean-square (RMS) energy. 
We evaluate two tasks: (i) \textbf{PSG-Audio (Binary)}, which distinguishes normal breathing from any apneic event across all segments (70,825 train / 21,349 test) using balanced accuracy; and (ii) \textbf{PSG-Audio (Subtype)}, which performs 4-class classification among pathological events (\emph{hypopnea}, \emph{mixed apnea}, \emph{obstructive apnea}, and \emph{central apnea}; 45,120 train / 13,765 test segments) using Macro-F1.

\item \textbf{CirCor Murmur.}
The CirCor DigiScope Phonocardiogram dataset~\citep{oliveira2022circor,oliveira2022circorrelease} comprises 3,163 heart-sound recordings from 942 pediatric patients, captured across up to four standard auscultation locations per patient and annotated for murmur presence (\emph{absent}, \emph{present}, or \emph{unknown}). We use a stratified patient-disjoint 80/20 split comprising 755 training patients (2,534 recordings) and 187 test patients (629 recordings). Because CirCor recordings are natively sampled at 4~kHz with diagnostic cardiac energy concentrated in 20-600~Hz, standard resampling to 16~kHz would compress the signal into the lowest bins of an 8~kHz Mel filterbank while leaving frequencies above 2~kHz empty. Instead, we feed the native 4~kHz samples directly to the 16~kHz frontend without interpolation ($4\times$ rate shift), mapping the 20-600~Hz cardiac band into 80-2,400~Hz where pretrained Mel filterbanks have high resolution. Each recording is encoded via overlapping 10.24-second windows (5.12-second stride) and mean-pooled into a recording-level representation. At test time, predictions across all auscultation locations of a patient are aggregated via the official PhysioNet clinical escalation rule (\emph{present} $>$ \emph{unknown} $>$ \emph{absent}) and evaluated by patient-level Macro-F1.
\end{itemize}

\paragraph{Speech Emotion Recognition.}\mbox{}\par
\label{sec:app-ser}
We evaluate frozen representations on three acted-speech corpora to examine whether self-supervised acoustic representations retain vocal affect and paralinguistic cues. CREMA-D and RAVDESS use \textbf{actor-disjoint} partitions to assess generalization to unseen speakers, while TESS uses a \textbf{word-disjoint} partition to assess generalization to unseen lexical content. 

\begin{itemize}[leftmargin=1.5em, labelsep=0.5em, itemsep=0.35em, topsep=0.3em, parsep=0pt, partopsep=0pt, beginpenalty=10000]
\item \textbf{CREMA-D.}
CREMA-D~\citep{cao2014crema} comprises audio-visual recordings of twelve sentences spoken by 91 actors (48 male, 43 female) of diverse ages and ethnic backgrounds across six core emotion categories: anger, disgust, fear, happiness, neutrality, and sadness. 
We exclude a single corrupted zero-energy recording, yielding 7,441 valid clips. 
We establish an actor-disjoint partition: 73 actors (5,965 clips) form the training set, while 18 unseen actors (1,476 clips) form the test set. 
The test cohort is strictly stratified to balance sex (9 male, 9 female) and demographic diversity. 
The prediction target is 6-class classification, evaluated by macro-averaged F1 (Macro-F1).

\item \textbf{RAVDESS.}
The Ryerson Audio-Visual Database of Emotional Speech and Song~\citep{livingstone2018ravdess} contains 1,440 speech recordings produced by 24 professional actors (12 male, 12 female) across eight emotions: neutral, calm, happy, sad, angry, fearful, disgust, and surprised. 
We employ the audio speech subset and construct an actor-disjoint split: Actors 01-20 (10 male, 10 female; 1,200 clips) serve as training data, while Actors 21-24 (2 male, 2 female; 240 clips) are held out exclusively for testing. 
The prediction target is 8-class classification, evaluated by Macro-F1.

\item \textbf{TESS.}
The Toronto Emotional Speech Set~\citep{pichora2020tess} contains 2,800 recordings from two actresses: a young adult female (YAF, 26 years old) and an older adult female (OAF, 64 years old). 
Each speaker utters 200 target words embedded in the carrier phrase \emph{``Say the word [target]''} across seven emotional states: anger, disgust, fear, happiness, pleasant surprise, sadness, and neutrality. 
For utterance-level probing, we use a \textbf{word-disjoint} partition over the 200 target words: 160 words (2,240 clips, 1,120 per speaker) form the training set, while 40 unseen words (560 clips, 280 per speaker) form the test set, fully balanced across both speakers and seven emotions. Both speakers appear in training and test data, so this protocol evaluates generalization to unseen words. Cross-speaker transfer is evaluated separately in \S\ref{sec:app-zeroshot-emotion}.
We evaluate 7-class classification using Macro-F1.
\end{itemize}

\subsubsection{Frame-Level Linear Probing for Sound Event Detection}
\label{sec:app-frame-probing}

\paragraph{AudioSet Strong and temporal targets.}
We use AudioSet Strong~\citep{hershey2021benefit} to assess whether frozen representations support temporally localized event predictions. Unlike utterance-level classification, this task retains the onset and offset annotations and predicts class activity along the recording. A linear classification layer operates on the frozen time-indexed representations to evaluate frame-level localization without encoder adaptation.

\paragraph{Frame-wise and event-based metrics.}
We report frame-wise mAP and PSDS1/PSDS2 to assess frame-level
classification and event detection, respectively.
For frame-wise mAP, we pool frame-level predictions and ground-truth
labels across all evaluation recordings, compute average precision
(AP) separately for each class, and take the unweighted mean across
classes.
For event detection, we report PSDS1 and PSDS2 under the Polyphonic
Sound Detection Score framework~\citep{bilen2020framework}, which
evaluates detected event intervals across operating points using
temporal-intersection criteria.

\FloatBarrier

\subsection{Zero-Shot Emotion Transfer: Protocol Details}
\label{sec:app-zeroshot-emotion}

This section details the source and target splits for the zero-shot emotion transfer evaluation in \S\ref{sec:exp-zeroshot}. 
All recordings are resampled to 16~kHz mono and symmetrically zero-padded to the 10.24-second input window. Strict linear probes are trained on frozen source-domain representations with cross-entropy loss and applied directly to the target domain without further training, reporting Macro-F1 (\%; mean $\pm$ standard deviation over three seeds).

\paragraph{Cross-corpus transfer (CREMA-D $\to$ RAVDESS / TESS).}
We directly apply the 6-class probe trained on the CREMA-D training split (\S\ref{sec:app-ser}) to all recordings of the six shared emotion classes in the target corpora: RAVDESS (excluding \emph{calm} and \emph{surprised}, leaving 1,056 clips) and TESS (excluding \emph{pleasant surprise}, leaving 2,400 clips).
These settings assess transfer across corpora with different speakers, lexical content, and recording conditions.

\paragraph{Cross-speaker transfer (TESS YAF $\leftrightarrow$ OAF).}
We train a 7-class linear probe on all clips of the younger speaker (YAF, 26 years old) and evaluate it directly on all clips of the older speaker (OAF, 64 years old), then reverse the direction. This isolates generalization across speaker age and vocal timbre under matched lexical content.

\subsection{Zero-Shot RealDESED Evaluation Protocol and Class Mapping}
\label{sec:app-desed-breakdown}

Table~\ref{tab:desed-class-mapping} specifies the ontology-based class mapping used for zero-shot cross-dataset sound event tagging on RealDESED (\S\ref{sec:exp-zeroshot}). We evaluate 12 of its 15 event classes, each mapped to one or more corresponding AudioSet categories, while excluding the remaining three classes (\textit{coffee machine}, \textit{light switch}, and \textit{window open/close}) that lack direct AudioSet counterparts.

\begin{table}[!htbp]
\centering
\small
\caption{\textbf{Ontological Mapping Between RealDESED Classes and AudioSet Categories} for cross-dataset sound event tagging.}
\label{tab:desed-class-mapping}
\begingroup
\setlength{\tabcolsep}{5pt}
\renewcommand{\arraystretch}{1.06}
\begin{tabularx}{\linewidth}{@{}llX@{}}
\toprule
\textbf{RealDESED Target Class} & \textbf{AudioSet Ontology ID} & \textbf{AudioSet Label Description} \\
\midrule
\texttt{bell\_ringing} & \texttt{/m/0395lw} & Bell \\
\texttt{cutlery\_dishes} & \texttt{/m/04brg2, /m/023pjk} & Dishes, pots, and pans; Cutlery, silverware \\
\texttt{door\_open\_close} & \texttt{/m/02dgv} & Door \\
\texttt{footsteps} & \texttt{/m/07pbtc8} & Walk, footsteps \\
\texttt{keyboard\_typing} & \texttt{/m/01m2v, /m/0316dw} & Computer keyboard; Typing \\
\texttt{keychain} & \texttt{/m/03v3yw} & Keys jangling \\
\texttt{microwave} & \texttt{/m/0fx9l} & Microwave oven \\
\texttt{phone\_ringing} & \texttt{/m/07pp8cl, /m/01hnzm} & Telephone bell ringing; Ringtone \\
\texttt{running\_water} & \texttt{/m/02jz0l} & Water tap, faucet \\
\texttt{toilet\_flushing} & \texttt{/m/01jt3m} & Toilet flush \\
\texttt{vacuum\_cleaner} & \texttt{/m/0d31p} & Vacuum cleaner \\
\texttt{wardrobe\_drawer\_open\_close} & \texttt{/m/0642b4, /m/0fqfqc} & Cupboard open or close; Drawer open or close \\
\bottomrule
\end{tabularx}
\endgroup
\end{table}

We evaluate all 1,008 recordings in the RealDESED test split (resampled to 16~kHz mono), collapsing the onset/offset event annotations into clip-level multi-hot ground-truth targets indicating whether each class occurs anywhere within a recording (retaining recordings without target events as all-negative examples). Because RealDESED recordings extend up to $\sim$31~seconds and exceed the 10.24-second input context of the pretrained encoders, each recording is partitioned into consecutive non-overlapping 10.24-second windows (with the final tail window zero-padded to 10.24~seconds). Each window is scored independently by the frozen encoder and its AudioSet-2M fine-tuned 527-way classification head, and the resulting 527-dimensional predictions are aggregated across all windows of a recording via element-wise max-pooling. Finally, for RealDESED classes mapped to multiple AudioSet categories in Table~\ref{tab:desed-class-mapping}, the target class score is obtained by taking the maximum score over its mapped AudioSet nodes, and we report macro-averaged mAP across the 12 classes.

\end{document}